\documentclass[twocolumn, showpacs,preprintnumbers,amsmath,amssymb]{revtex4}

\usepackage{graphicx}

\begin{document}

\addtolength{\textheight}{1.2cm}
\addtolength{\topmargin}{-0.5cm}

\newcommand{\etal} {{\it et al.}}

\title{Energy as witness of multipartite entanglement in spin clusters}

\author{F. Troiani$^1$ and I. Siloi$^{1,2}$}

\affiliation{$^{1}$ S3 Istituto Nanoscienze-CNR, Modena, Italy}
\affiliation{$^{2}$ Dipartimento di Fisica, Universit\`a di Modena e Reggio Emilia, 
Italy}

\date{\today}

\begin{abstract}

We develop a general approach for deriving the energy minima of biseparable states in chains of arbitrary spins $s$, and report numerical results for spin values $s \le 5/2$ (with $N \le 8$). 
The minima provide a set of threshold values for exchange energy, that allow to detect different degrees of multipartite entanglement in one-dimensional spin systems.
We finally demonstrate that the Heisenberg exchange Hamiltonian of $N$ spins has a nondegenerate $N$-partite entangled ground state, and can thus witness such correlations in all finite spin chains.

\end{abstract}


\maketitle

Entanglement is one of the most striking peculiarities of quantum systems and promises to play a crucial role in emerging quantum technologies \cite{amico}.
This has fueled the development of theoretical and experimental means for its detection in 
diverse physical systems \cite{guhne09}.
One of the most convenient such tools is represented by entanglement witnesses. These are
observables whose expectation value can exceed given bounds only in the presence of specific forms of entanglement. 
Macroscopic observables such as magnetic susceptibility \cite{wiesniak05,brukner06,ghosh} 
and internal 
energy \cite{brukner04,toth05,PhysRevA.70.062113}
allow for example to discriminate between fully separable and entangled spin states.
In qubit systems, further inequalities for energy have been derived, whose violation 
implies multipartite entanglement \cite{guhne05,PhysRevA.73.052319}.
Along the same lines, the measurement of collective observables \cite{PhysRevLett.107.180502} 
allows to detect multipartite entanglement in the vicinity of prototypical quantum states through
the spin-squeezing inequalities \cite{sorensen,PhysRevLett.95.120502}.
The connection of these studies with quantum-information processing has however focused most of the attention on entanglement between qubits. 
Limited attention has instead been devoted to multipartite entanglement between composite systems of $s>1/2$ 
(pseudo)spins \cite{PhysRevLett.107.240502}.

In the present paper, we address the problem of detecting multipartite entanglement 
in clusters of arbitrary spins $s$ \cite{wangp,brus} using exchange energy as a witness.
We develop a general approach for deriving the energy minima $ \bar E_{bs}^N $ of biseparable 
states $ | \psi_A \rangle \otimes | \psi_B \rangle $ in chains of $N$ spins, that exploits 
the rotational symmetry of the system Hamiltonian. This allows to reduce the minimization 
problem to calculating the ground states of effective spin Hamiltonians within each 
subsystem $A$ and $B$. 
Minima derived for $k-$spin chains provide in turn a set of threshold values for energy, corresponding to $k$-partite entanglement in chains of $n_k(k-1)+1$ or rings 
of $n_k(k-1)$ spins. 
Analytical expressions of the minima are derived for the simplest cases, while numerical 
solutions are provided for $ s \le 5/2 $, that correspond to prototypical 
models of molecular nanomagnets \cite{candini10,troiani11,troiani05,trif08}.
As a general result, we finally demonstrate that the ground state of an $N$-spin chain with Heisenberg Hamiltonian is $N$-partite entangled.
This implies an energy gap between biseparable and $N$-spin entangled states, and the 
possibility of detecting the latter ones by exchange energy, in finite spin chains with 
arbitrary $N$ and $s$.

\begin{table*}
\vspace{0.5cm}
\begin{tabular}{c|ccccccccccc}
\hline \hline
$s$& \ \ \ $\bar a_0$ \ \ \ 
& \ \ \ $\bar a_1$ \ \ \ 
& \ \ \ $\bar a_2$ \ \ \ 
& \ \ \ $\bar a_3$ \ \ \ 
& \ \ \ $\bar a_4$ \ \ \ 
& \ \ \ $\bar a_5$ \ \ \ 
& $\bar \alpha_{S+1} - \bar \alpha_{S} $ 
& $ \bar E_{bs}^{3} $ & $\bar E_{12}$ & $\bar E_{23} $ & $E_0$ \\
\hline 
$ 1/2 $ & 0.973 & 0.230 &      - &      - &       - &       - & $ \pi   $ & -0.8090 & -0.1118 & -0.6972 & - 1.0 \\
$ 1   $ & 0.858 & 0.506 & 0.0839 &      - &       - &       - & $ \pi   $ & -2.481  & -0.7583 & -1.722  & - 3.0 \\
$ 3/2 $ & 0.749 & 0.631 & 0.198  & 0.0269 &       - &       - & $ \pi   $ & -5.162  & -1.933  & -3.230  & - 6.0 \\
$ 2   $ & 0.671 & 0.676 & 0.298  & 0.0696 & 0.00819 &       - & $ \pi $ & -8.849  & -3.601  & -5.248  & -10.0 \\
$ 5/2 $ & 0.612 & 0.687 & 0.373  & 0.120  & 0.0232  & 0.00244 & $ \pi $ & -13.54  & -5.768  & -7.771  & -15.0 \\
\hline \hline
\end{tabular}
\caption{\label{table1} Minima $\bar E_{bs}^{3}$ of $H_{123}$ for biseparable states 
$ |\psi_{1} \rangle \otimes |\psi_{23} \rangle $ and corresponding coefficients $\bar a_S$ and $\bar \alpha_S $. 
The ground state energies $E_0$ of $H_{123}$ are also reported, as well as 
$\bar E_{12}=-s f(\bar{\bf a},\bar{\bf \alpha}) $
and
$\bar E_{23}=   g(\bar{\bf a},\bar{\bf \alpha}) $. 
} 
\end{table*}
{\it Tripartite entanglement ---}
Tripartite entangled states are detected by a three-spin Hamiltonian 
$ H_{123} $  if their energy exceeds the lower bound that applies to biseparable states \cite{guhne05}. 
Here, we seek such bound for 
$
H_{123} = {\bf s}_1 \cdot {\bf s}_2 + {\bf s}_2 \cdot {\bf s}_3  \equiv H_{12} + H_{23}
$,
and for a generic biseparable state
$ 
| \psi_1 \rangle \otimes | \psi_{23} \rangle 
$:
\begin{equation}\label{eq01}
\bar E_{bs}^{3}\!\! =\!\!\!\! \min_{|\psi_1\rangle, |\psi_{23}\rangle} \!\!\left\{
\langle \psi_1 | {\bf s}_1 | \psi_1 \rangle\! \cdot \!
  \langle \psi_{23} | {\bf s}_2 | \psi_{23} \rangle\!\! +\!
    \langle  \psi_{23} | {\bf s}_2\!\cdot\! {\bf s}_3  | \psi_{23} \rangle \right\}\! .
\end{equation}
If we identify the direction of 
$ \langle \psi_{23} | {\bf s}_2 | \psi_{23} \rangle $ with the $z$ axis,
the first term in Eq. \ref{eq01} simplifies to:
$
\langle H_{12} \rangle = \langle \psi_1 | s_{1,z} | \psi_1 \rangle \langle \psi_{23} | s_{2,z} | \psi_{23} \rangle 
$,
where $ \langle \psi_{23} | s_{2,z} | \psi_{23} \rangle \ge 0 $ by definition.
For any given $ | \psi_{23} \rangle $, the state of $ s_1 $ that minimizes 
$ \langle H_{123} \rangle $ is thus given by
$
| m_1\!\! =\!\! - s_1 \rangle 
$,
and the problem of deriving $ \bar E_{bs}^3 $ reduces to finding 
the state $ | \psi_{23} \rangle $ 
that minimizes:
\begin{equation}\label{eq02}
\langle  \psi_{23} | \tilde{H}_{23} | \psi_{23} \rangle
\equiv 
\langle \psi_{23} | -s_1 s_{2,z} + {\bf s}_2 \cdot {\bf s}_3  | \psi_{23} \rangle ,
\end{equation}
i.e. the ground state of the two-spin Hamiltonian $\tilde{H}_{23}$.
In order to derive the energy minima, it is convenient to 
expand $ | \psi_{23} \rangle $ in the form:
\begin{equation}\label{eq03}
| \psi_{23} \rangle = \!\!\!\!\!\!\sum_{M=-s_2-s_3}^{s_2+s_3}\!\!\!\!\!\!  \sqrt{P_M} \!\!\!\sum_{S=|M|}^{s_2+s_3}\!\!\! A^M_S |S,M\rangle 
\equiv \!\!\!\!\!\!\sum_{M=-s_2-s_3}^{s_2+s_3}\!\!\!\!\!\!  \sqrt{P_M} |\psi_{23}^M\rangle ,
\end{equation}
where $ {\bf S} = {\bf s}_2 + {\bf s}_3 $ and $M$ is its projection 
along $z$. 
Each real coefficient $ P_M $ gives the probability that ${\bf S}$ has 
a $z-$projection $M$ ($ \sum_M P_M = 1$). The normalization condition for the complex coefficients $ A^M_S = a^M_S e^{i\alpha _M^S} $ reads: $ \sum_S (a^M_S)^2 = 1$ (with $ a^M_S = |A^M_S| $).
Given that both the operators $s_{2,z}$ and ${\bf s}_2 \cdot {\bf s}_3 $ commute with 
$S_z$, the energy expectation value can be written as
$
\langle \tilde{H}_{23} \rangle  = 
\sum_M P_M E_{bs}^{3,M} 
$,
where,
\begin{equation}\label{eq04}
E_{bs}^{3,M}\!\! =\! 
\langle \psi_{23}^M | \tilde{H}_{23} | \psi_{23}^M \rangle  
\!\equiv\!
\!-\!s_1 f_M({\bf a}^M\!\!,{\bf \alpha}^M\!)\!+\!g_M({\bf a}^M\!\!,{\bf \alpha}^M\!),
\end{equation}
with ${\bf a}^M\!\! =\!\! (a_{|M|}^M, \dots , a^M_{s_i+s_j}) $
and  ${\bf \alpha}^M\!\! =\!\! (\alpha_{|M|}^M, \dots , \alpha^M_{s_i+s_j}) $.
The energy expectation value is thus given by an average, with probabilities $P_M$, of functions $ E_{bs}^{3,M} $ that depend on disjoint groups 
of variables $ {\bf A}^M$, each corresponding to a given 
$M$. 
This allows to minimize the terms $ E_{bs}^{3,M} $ independently from one another, and to 
identify the overall minimum with the lowest $ \bar E_{bs}^{3,M} $:
\begin{equation}
\bar E_{bs}^{3} = 
\min_M \bar E_{bs}^{3,M} ( \bar {\bf a}^M, \bar{\bf\alpha}^M ) .
\end{equation}
The dependence of $E_{bs}^{3,M}$ on the variables $A_S^M$ is derived as follows.
The first contribution in Eq. \ref{eq04} is proportional to:
$
f_M = \langle s_{2,z} \rangle =
\sum_{S,S'} (A_S^M)^* (A_{S'}^M) \langle S,M | s_{2,z} | S', M \rangle 
$.
Here, the matrix element can be expressed in terms of the Clebsch-Gordan coefficients 
\cite{tsukerblat}:
$
\langle S,M | s_{2,z} | S', M \rangle = 
\sum_{m_2} \langle S,M | m_2 , m_3 \rangle\langle m_2 , m_3 | S', M \rangle m_2
$ (with $m_3 = M-m_2$).
The second contribution in Eq. \ref{eq04} is instead diagonal in the basis 
$ | S , M \rangle $, and reads:
$
g_M = \langle {\bf s}_2 \cdot {\bf s}_3 \rangle = 
\sum_S (a^M_S)^2 [S(S+1)-s_2(s_2+1)-s_3(s_3+1)] / 2
$.

In order to analytically minimize - for $s \le 3/2$ - the function $ E_{bs}^{3,M} $ subject to the normalization 
constraints, we apply the method of Lagrange 
multipliers.
The stationary points of the Lagrange function 
$
\Lambda_M (A^M_S,\lambda) = 
E_{bs}^{3,M} + \lambda [\sum_S (a^M_S)^2-1] 
$,
are identified by the equations
$ 
\partial \Lambda_M / \partial a^M_S = 
  \partial \Lambda_M / \partial \alpha^M_S = 
  \partial \Lambda_M / \partial \lambda = 0 
$,
for $ |M| \le S \le s_2+s_3 $.
In all the cases considered below, the lowest minima correspond to $M=0$: $ \bar E^3_{bs} = \bar E_{bs}^{3,M=0} $.
We shall thus refer only to this subspace, and omit
the apices $M$ from the notation. Besides, we focus on the case 
of identical spins.

In the $s=1/2$ case, a lower bound for $ \langle H_{123} \rangle $ in the 
absence of tripartite entanglement has already been derived by different means 
\cite{guhne05}.
Here we show that such value actually corresponds to a minimum, and derive the 
corresponding biseparable state.
The dependence of $ E_{bs}^{3} $ on the parameters $a_S$ and $\alpha_S$ is given by (see Eq. \ref{eq04}):
$
f_M = - a_0a_1 \cos (\alpha_0-\alpha_1)  
$
and
$ g_M = (-3a_0^2+a_1^2)/4 $.
As far as the phases $ \alpha_S $ are concerned, $ E^3_{bs} $ is minimized by 
$\bar\alpha_1 - \bar\alpha_0 = \pi $.
The remaining conditions give rise to the energy minimum 
$ \bar E_{bs}^{3} = -(1+\sqrt{5})/4 $,
that coincides with the lower bound derived in Ref. \cite{guhne05}. The corresponding 
biseparable state is  given by:
\begin{equation}
\bar a_0 = \left( 1/2 + 1/\sqrt{5} \right)^{1/2}, \ 
\bar a_1 = \left( 1/2 - 1/\sqrt{5} \right)^{1/2}.
\end{equation}

We proceed in the same way in the case $s=1$, 
where the expression of energy is given by:
$
f_M = 2\, a_1 (a_0\sqrt{2}+a_2) /\sqrt{3} 
$
and
$
g_M =  -2a_0^2-a_1^2+a_2^2
$.
Here, the conditions
$\bar\alpha_{S+1} - \bar\alpha_S = \pi $,
derived from 
$ \partial \Lambda_M / \partial \alpha_S = 0 $,
have already been included.
The analytic expression of the energy minimum is:
\begin{equation}\label{eqe3}
\bar E_{bs}^{3}\!\! =\! -2/3\! \left\{ 1\!\! +\! \sqrt{5/2}\! \left[ \cos (\varphi /3) + \sqrt{3} \sin (\varphi /3) \right] \right\},
\end{equation}
where 
$ \varphi = \arccos [1 / (10\sqrt{10})] $.

For the spin values $s=3/2$, $s=2$, and $s=5/2$, we directly report 
the energy minima, and the corresponding biseparable states (Table \ref{table1}),
that have been obtained through a conjugate gradient algorithm \cite{atkinson}.

The comparison between the different spin values shows that the relative weight of 
the singlet state ($\bar a_0$) decreases with increasing $s$, as well as the ratio 
between the energies of the entangled and unentangled spin pairs 
($ \bar E_{23} / \bar E_{12} $).
In all cases, the inequality
$ \langle H_{123} \rangle < \bar E_{bs}^{3} $
implies tripartite entanglement in the three-spin system.
The criterion becomes 
$ \langle H \rangle < n_3 \bar E_{bs}^{3} $
for any $H$ that can be written as the sum of $n_3$ three-spin Hamiltonians,
such as chains of $2n_3+1$ spins or rings with $2n_3$. 

{\it Quadripartite entanglement ---}
We consider the expectation values of the four-spin Hamiltonian
$
H_{1234} = {\bf s}_1 \cdot {\bf s}_2 
         + {\bf s}_2 \cdot {\bf s}_3
         + {\bf s}_3 \cdot {\bf s}_4
$,
corresponding to the biseparable states
$
| \psi_{22}^4  \rangle = | \psi_{12} \rangle \otimes | \psi_{34} \rangle
$
and
$
| \psi_{13}^4 \rangle = | \psi_{1} \rangle \otimes | \psi_{234} \rangle
$.
In the former case, we compute:
\begin{eqnarray}
\bar E_{22}^4
        &=& \min_{| \psi_{12} \rangle , | \psi_{34} \rangle} \left\{
           \langle \psi_{12} | {\bf s}_1 \cdot {\bf s}_2 | \psi_{12} \rangle 
         + \langle \psi_{34} | {\bf s}_3 \cdot {\bf s}_4 | \psi_{34} \rangle
\right. \nonumber\\ 
        &+&\left.            
           \langle \psi_{12} | s_{2,z}   | \psi_{12} \rangle  
           \langle \psi_{34} | s_{3,z}   | \psi_{34} \rangle
            \right\} ,
\end{eqnarray}
where $z$ is defined as the direction of 
$ \langle \psi_{34} | s_{3,z}   | \psi_{34} \rangle $.
The states $ | \psi_{12} \rangle $ and $ | \psi_{34} \rangle $ are expanded in the 
bases $ | S=S_{12} , M=M_{12} \rangle $ and  $ | S'=S_{43} , M'=M_{43} \rangle $, respectively.
For $ | \psi_{12} \rangle $, we use the expression in Eq. \ref{eq03} (and  
replace the indices $23$ with $12$).
Similarly, $ | \psi_{34} \rangle $ is expressed as:
$
| \psi_{34} \rangle = \sum_{M'} \sqrt{Q_{M'}} \sum_S B^{M'}_{S'} |S',M'\rangle ,
$
with
$ B^{M'}_{S'} = b^{M'}_{S'} e^{i \beta^{M'}_{S'}} $, 
$ \sum_{M'} Q_{M'} = \sum_{S'} (b^{M'}_{S'})^2 = 1 $.
Being both $M$ and $M'$ good quantum numbers, 
$
E_{22}^4 = \sum_M \sum_{M'} P_M Q_{M'} E_{22}^{4,MM'} 
$,
where:
\begin{equation}
E_{22}^{4,MM'}\!\! = g_M   ({\bf A}^M   ) 
  \!+\! f_M({\bf A}^M) f_{M'}({\bf B}^{M'}) 
                \!+\! g_{M'}({\bf B}^{M'}) 
\end{equation}
and the functions $f_M$ and $g_M$ coincide with those reported in 
the previous section.
\begin{table*}
\vspace{0.5cm}
\begin{tabular}{c|cccccccc|cccc}
\hline \hline
$s$ & $\bar a_0$ 
& $\bar a_1$ 
& $\bar a_2$ 
& $\bar a_3$ 
& $\bar a_4$ 
& $\bar a_5$ 
& $ \bar E_{bs}^4 \!=\!\bar E_{22}^4 $ 
& $ \bar E_{13}^4 $ 
& $\bar E_{bs}^{5} $
& $\bar E_{bs}^{6} $
& $\bar E_{bs}^{7} $
& $\bar E_{bs}^{8} $\\
\hline 
$ 1/2 $ & \ 1     & 0     & -      & -      & -       & - 
& - 1.500 & - 1.190\ \   
&\  -1.780 & -2.366 &  -2.697 & -3.244 \\
$ 1   $ & \ 0.921 & 0.387 & 0.0418 & -      & -       & - 
& - 4.051 & - 3.828\ \ 
&\  -5.343 & -6.771 & -8.133 & -9.537 \\
$ 3/2 $ & \ 0.775 & 0.607 & 0.171  & 0.0281 & -       & - 
& - 8.131 & - 7.957\ \   
&\  -10.90 & -13.75 & -16.56 & -19.39 \\
$ 2   $ & \ 0.687 & 0.669 & 0.278  & 0.0602 & 0.00649 & -   
& -13.74  & -13.59\ \   
&\  -18.46 & -23.24 & -27.99 & -32.75 \\
$ 5/2 $ & \ 0.627 & 0.669 & 0.359  & 0.134  & 0.110   & $<10^{-4}$ 
& -21.18  & -20.71\ \  
&\  -28.03 & -35.23 & -42.42 & -49.62 \\
\hline \hline
\end{tabular}
\caption{Left: Minima $\bar E_{22}^4$ and $\bar E_{13}^4$ for biseparable four-spin states 
$ |\psi_{12} \rangle \otimes |\psi_{34} \rangle $ and
$ |\psi_{1} \rangle \otimes |\psi_{234} \rangle $,
respectively.
The states corresponding to the former partition are given by the displayed values of $ \bar a_S $,
and by: $ \bar {\bf b}_S = \bar {\bf a}_S $, 
$ \bar \alpha_{S+1} - \bar \alpha_{S} = \pi$,
and 
$ \bar \beta_{S'+1} - \bar \beta_{S'} = 0 $.
Right: Energy minima $ \bar E_{bs}^N $ of $N$-spin systems.}
\label{table2}
\end{table*}
The energy $ E_{22}^{4,MM'} $ is minimized numerically by the conjugate gradient 
approach as a function of ${\bf a}^M$ and ${\bf b}^M$, while the 
minimization with respect to ${\bf \alpha}^M$ and ${\bf \beta}^M$ is straightforward.
The minimum $\bar E^4_{22}$ is then identified with the lowest 
$\bar E^{4MM'}_{22}$:
\begin{equation}
\bar E^4_{22} = 
\min_{M,M'} \bar E_{22}^{4,MM'} 
( \bar {\bf a}^M , \bar {\bf \alpha}^M, \bar{\bf b}^{M'} , \bar{\bf \beta}^{M'} ) .
\end{equation}
For all values of $s$, the lowest minima belong to the subspace $M=M'=0$.
The minimum of $\bar E^4_{13}$ is instead identified with the ground state energy 
of the three-spin Hamiltonian 
$ \tilde H_{234} = -s_1 s_{2,z} + {\bf s}_2 \cdot {\bf s}_3 + {\bf s}_3 \cdot {\bf s}_4 $, 
which belongs,  
in all the considered cases, to the subspace with  $ M = s $.
The energy minima and the corresponding states are reported in the left part of  
Table \ref{table2}. 
We note that the bipartition
$
| \psi_{12} \rangle \otimes | \psi_{34} \rangle
$
always gives lower minima with respect to
$
| \psi_{13}^4 \rangle = | \psi_{1} \rangle \otimes | \psi_{234} \rangle
$:
therefore, $ \bar E^4_{bs} = \bar E_{22}^4 $.
For the four-qubit system, the expectation value of energy is minimized 
by the dimerized state \cite{guhne05}. 
This is not the case for $s>1/2$, where the coupling between 
$s_2$ and $s_3$ induces a significant admixture with states of higher $S$ and $S'$.
Besides, the energy is minimized by the state 
with
$ \langle s_{2,z} \rangle = - \langle s_{3,z} \rangle $
($ \bar {\bf a}^0 = \bar {\bf b}^0 $ 
and 
$ \bar \beta_{S'+1}^0 - \bar \beta_{S'}^0 = \bar \alpha_{S+1}^0 - \bar \alpha_{S}^0 - \pi $).
We thus conclude that,
for all the considered spin values, the inequality 
$ \langle H_{1234} \rangle < \bar E_{bs}^4 $
implies quadripartite entanglement in the four-spin system.
The criterion generalizes to $ \langle H \rangle < n_4 \bar E_{bs}^4 $
for any $H$ that can be written as the sum of $n_4$ 
four-spin Hamiltonians, such as chains of $3n_4+1$ spins or rings with $3n_4$.

{\it N-partite entanglement ---}
For larger spin numbers $N$, the analytic derivation of the functions $f_M$ and $g_M$
becomes cumbersome, and a fully numerical approach is preferable.
Given a partition of the spin chain in two subsystems, $A$ and $B$, consisting of $N_A$ 
and $N_B=N-N_A$ consecutive spins, the Hamiltonian can be written as 
$ H = H_A + H_B + H_{AB} $,
where 
$ H_A = \sum_{i=    1}^{N_A-1} {\bf s}_i \cdot {\bf s}_{i+1} $,
$ H_B = \sum_{i=N_A+1}^{N  -1} {\bf s}_i \cdot {\bf s}_{i+1} $, 
and 
$ H_{AB} = {\bf s}_{N_A} \cdot {\bf s}_{N_A+1} $.
The energy minima for biseparable states
$ | \psi \rangle = | \psi_A \rangle \otimes | \psi_B \rangle $
are:
\begin{eqnarray}\label{eq09}
\bar E^{N}_{N_AN_B}\!\!\!\! & = & \!\!\!\!\!\!\min_{|\psi_A\rangle, |\psi_B\rangle} \!\!\!\left\{
     \langle \psi_A | H_A | \psi_A \rangle\!\! +\!
     \langle \psi_B | H_B | \psi_B \rangle\!\! 
\right. \nonumber\\ & + & \left.
     \langle \psi_A | {\bf s}_{N_A  }\! | \psi_A  \!\rangle \cdot 
     \langle \psi_B | {\bf s}_{N_A\!+\!1} | \psi_B  \!\rangle \right\},
\end{eqnarray}
We identify the $z$ direction with that of 
$ \langle \psi_A | {\bf s}_{N_A} | \psi_A  \rangle $ 
and define:
$
z_A\! \equiv \! \langle \psi_A | s_{N_A  ,z} | \psi_A  \rangle \ge 0 
$,
$ 
z_B\! \equiv \! \langle \psi_B | s_{N_A+1,z} | \psi_B  \rangle 
$.
Besides, the state $ | \bar\psi_B \rangle $ that minimizes $E_{bs}^{N_A,N_B}$ 
necessarily has an expectation value $ \langle {\bf s}_{N_A+1} \rangle $
antiparallel to $ \hat{\bf z} $ 
(and thus $z_B \le 0$): any 
rotation of the subsystem $B$ with respect to such orientation 
would in fact increase 
$ \langle H_{AB} \rangle $,
while leaving unaffected $\langle H_A + H_B \rangle$.
The minimization can now be split into two correlated eigenvalue problems, 
that consist in finding the ground states of 
$
\tilde{H}_A (z_B)\! =\! H_A\! +\! z_B s_{N_A  ,z}
$
and
$
\tilde{H}_B (z_A)\! =\! H_B\! +\! z_A s_{N_A+1,z}
$.
The self-consistent solution of the minimization problem Eq. \ref{eq09} is thus 
represented by the state
$ | \bar \psi \rangle = | \psi^0_A (\bar z_B) \rangle \otimes 
                        | \psi^0_B (\bar z_A) \rangle $
with:
$
\bar z_A = \langle \psi^0_A (\bar z_B) | s_{N_A  ,z} | \psi^0_A (\bar z_B) \rangle 
$
and
$
\bar z_B = \langle \psi^0_B (\bar z_A) | s_{N_A+1,z} | \psi^0_B (\bar z_A) \rangle 
$,
where
$ | \psi^0_A (\bar z_B) \rangle $
is the ground state of 
$ \tilde{H}_A (\bar z_B) $
and 
$ | \psi^0_B (\bar z_A) \rangle $
that of 
$ \tilde{H}_B (\bar z_A) $.
The corresponding value of energy is given by
\begin{equation}
\bar E^{N}_{N_AN_B} = E^0_A (\bar z_B) + E^0_B (\bar z_A) - \bar z_A \bar z_B ,
\end{equation}
where the last term avoids the double counting of the  
contribution from $ {\bf s}_{N_A} \cdot {\bf s}_{N_A+1}$.
The values of the overall minima for biseparable states, given by
\begin{equation}
\bar E^N_{bs} = \min_{N_A,N_B} \bar E_{N}^{N_AN_B} ,
\end{equation}
are reported in the right part of Table \ref{table2} for $N \le 8$.
For all the considered values of $s$ and $N$, the partition with lowest energy minimum
is that with $N_A=2$.
We note that for even $N_A$ and $N_B$, the qubits only present a solution 
with $ \langle H_{AB} \rangle = 0$; for $s>1/2$, instead, 
the minimum corresponds to the additional solution, with finite $ \langle H_{AB} \rangle $. As in the cases of tri- and quadri-partite entanglement, these minima provide 
a criterion, namely $\langle H \rangle < n_k \bar E^k_{bs} $, for the detection of 
$k$-partite entanglement in chains and rings with $n_k (k-1)+1$ and $n_k (k-1)$ spins,
respectively.

We finally demonstrate the presence of $N$-partite entanglement in the ground state of  
all spin chains with even $N$.

{\it Theorem. ---} The ground state $|\psi_0\rangle$ of the spin Hamiltonian 
$ H = \sum_{i=1}^{N-1} {\bf s}_i \cdot {\bf s}_{i+1}$, with even $N$, cannot be written 
in any biseparable form $|\psi_{AB}\rangle = |\psi_A\rangle \otimes |\psi_B\rangle$, 
and is thus $N$-partite entangled. 

{\it Proof. ---} 
According to Marshall's theorems \cite{auerbach}, $|\psi_0\rangle$ is a nondegenerate 
singlet state. 

A biseparable state $ |\psi_{AB}\rangle $ can only be a singlet if $S_\chi=0$ ($ \chi = A,B $). 
In fact, one can write $ |\psi_\chi\rangle $  as a linear superposition of eigenstates of ${\bf S}_\chi^2$: 
$|\psi_\chi\rangle = \sum_{S_\chi} C_{S_\chi}^\chi | \phi^\chi_{S_\chi} \rangle $.
The following inequality applies:
$ \langle {\bf S}^2\rangle 
\ge
\sum_{S_A,S_B} |C_{S_A}^A C^B_{S_B}|^2 [ (S_A - S_B)^2 + S_A + S_B ] 
\ge
\sum_{S_A,S_B} |C_{S_A}^A C^B_{S_B}|^2 ( S_A + S_B )$,
where we make use of:
$ \langle \phi^A_{S_A} | {\bf S}_A | \phi^A_{S_A} \rangle \cdot 
  \langle \phi^B_{S_B} | {\bf S}_B | \phi^B_{S_B} \rangle \ge 
- S_A S_B $. 
Therefore, $ \langle {\bf S}^2\rangle $ can only vanish 
$C^A_{S_A}=\delta_{S_A,0}$ and $C^B_{S_B}=\delta_{S_B,0}$.

We now prove that the state $ |\psi_A\rangle \otimes |\psi_B\rangle $, with 
$S_{A/B}=0$, cannot be the 
ground state of $H$, by showing that $ H | \psi_{AB} \rangle $ has a component 
$ | \psi_{AB}^\perp \rangle $
which is
orthogonal to $ | \psi_{AB} \rangle $.
To this aim, we write: 
$ H_{AB} = s_{z,N_A} s_{z,N_A+1} + 
(s_{+,N_A} s_{-,N_A+1} + {\rm h.c.} )/2$.
We first show that 
$ | \psi_{AB}^\perp \rangle \equiv  s_{z,N_A} s_{z,N_A+1} | \psi_{AB} \rangle $ is finite 
and belongs to the 
subspace $S_{A/B}=1$ and $M_{A/B}=0$.
In the partial spin sum basis \cite{tsukerblat}, the state of $A$ reads:
$ | \psi_A \rangle = \sum_\alpha D_\alpha | \alpha , S_{A} , M_{A} \rangle $,
where $ \alpha $ denotes the quantum numbers $ S_1, \dots , S_{N_A-1} $ corresponding 
to the partial spin sums ${\bf S}_k \equiv \sum_{i=1}^k {\bf s}_k$,
and $S_A=0$ implies $S_{N_A-1}=s$. 
The operator $ s_{N_A}^z $ commutes 
with all ${\bf S}_k^2$ with $ k \le N_A -1$.
The matrix elements of the $N_A-$th spin can thus be reduced to 
those between the states of two spins $s$: 
$ \langle \alpha' , S_{A}' , M_{A}' | s_{N_A}^z | \alpha , S_{A} , 
M_{A} \rangle 
= \delta_{\alpha , \alpha'} 
\langle S_{A}' , M_{A}' | s_{N_A}^z | 0 , 0 \rangle $. 
The latter matrix element is only finite, and equals $-\eta_s$, for 
$ S_A' =1$ and $ M_A' = 0$;
therefore,
$ s_{N_A}^z | \psi_A \rangle = - \eta_s
\sum_\alpha D^A_\alpha | \alpha , 1 , 0 \rangle $,
with 
$ \eta_s = [ ( \sum_{m=-s}^s m^2 ) / (2s+1)]^{1/2} > 0 $.
The same procedure can be applied to $B$, resulting in:
$ s_{N_A+1}^z | \psi_B \rangle = - \eta_s
\sum_\beta D^B_\beta | \beta , 1 , 0 \rangle $.
Here 
$ | \psi_B \rangle = \sum_\beta D^B_\beta | \beta , S_B=0 , M_B = 0 \rangle $,
and 
$ \beta $ denotes the quantum numbers $ S_1 , \dots , S_{N_B-1} $ 
corresponding to $ {\bf S}_k = \sum_{i=1}^k {\bf s}_{N+1-i} $.
As a result, 
$ | \psi_{AB}^\perp \rangle  
= \eta_s^2
\sum_{\alpha ,\beta} D^A_\alpha D^B_\beta | \alpha , 1 , 0 \rangle \otimes 
| \beta  , 1 , 0 \rangle $ has finite norm, belongs to the subspace $S_{A/B}=1$ and 
$M_{A/B}=0$, and is thus orthogonal to $ | \psi_{AB} \rangle $. 

We now show that $ | \psi_{AB}^\perp \rangle $ coincides with the component of 
$H | \psi_{AB} \rangle $ with $S_{A/B}=1$ and $M_{A/B}=0$.
In fact, 
$ (H_A+H_B) | \psi_{AB} \rangle $
belongs to the $S_{A/B}=0$ subspace,
being
$ [H_\chi,{\bf S}^2_{\chi'}] = 0 $ for $\chi, \chi' = A,B$.
The states
$ s_{N_A}^\pm s_{N_A+1}^\mp | \psi_{AB} \rangle $
belong instead to the subspaces 
$ M_A = - M_B = \pm 1 $.
Therefore, $ H| \psi_{AB} \rangle $ has a finite component $ | \psi^\perp_{AB}\rangle $,
and cannot be an eigenstate of $H$.

We finally consider the case where the spins of the subsystems are not consecutive.  
In the simplest case, the spins of $A$ are split into two sequences 
of $N_{A_1}$ and $N_{A_2}$ consecutive spins, separated by the $N_B$ spins of $B$.
If $ | \psi_A \rangle = | \psi_{A_1} \rangle \otimes | \psi_{A_2} \rangle $, then 
this case can be recast into the previous one, by redefining $A'=A_1$ and 
$B'=B \cup A_2$.
If instead $A_1$ and $A_2$ are entangled, 
then 
$
| \psi_{AB} \rangle
$,
is degenerate with any 
$
| \psi_{AB}' \rangle 
$,
where $|\psi_A\rangle $ is replaced by a state $|\psi_A'\rangle $
that gives the same reduced density matrices $ \rho_{A_k} $ for $A_1$ and $A_2$; 
this is because correlations 
between uncoupled spins don't affect $ \langle H \rangle $.
Therefore, the state of $ | \psi_{AB} \rangle $ would be degenerate, which contradicts 
Marshall's theorems. 
The same conclusion can be drawn for any bipartition 
where $A$ and $B$ don't consist of consecutive spins, by recursively applying the 
above argument.  
$ \blacksquare $

In conclusion, we have developed a simple approach for deriving the energy minima of biseparable states in chains of arbitrary spins $s$. These minima can be used for detecting $k$-partite entanglement in chains with $n_k(k-1)+1$ and rings with $n_k(n-1)$ spins, respectively. 
This approach has been here applied to spin chains of up to $8$ spins $s \le 5/2$. Finally, we have demonstrated on 
general grounds that the Heisenberg interaction induces $N$ partite entanglement in the nondegenerate ground state of even-numbered chains with arbitrary $s$. Such entanglement 
can thus always be detected by using energy as a witness.

We thank F. Manghi, M. Rontani, and A. Bertoni for useful discussions. 
We acknowledge financial support from PRIN of the Italian MIUR.


\begin{thebibliography}{23}

\expandafter\ifx\csname natexlab\endcsname\relax\def\natexlab#1{#1}\fi
\expandafter\ifx\csname bibnamefont\endcsname\relax
  \def\bibnamefont#1{#1}\fi
\expandafter\ifx\csname bibfnamefont\endcsname\relax
  \def\bibfnamefont#1{#1}\fi
\expandafter\ifx\csname citenamefont\endcsname\relax
  \def\citenamefont#1{#1}\fi
\expandafter\ifx\csname url\endcsname\relax
  \def\url#1{\texttt{#1}}\fi
\expandafter\ifx\csname urlprefix\endcsname\relax\def\urlprefix{URL }\fi
\providecommand{\bibinfo}[2]{#2}
\providecommand{\eprint}[2][]{\url{#2}}

\bibitem[{\citenamefont{Amico et~al.}(2008)\citenamefont{Amico, Fazio,
  Osterloh, and Vedral}}]{amico}
\bibinfo{author}{\bibfnamefont{L.}~\bibnamefont{Amico}},
  \bibinfo{author}{\bibfnamefont{R.}~\bibnamefont{Fazio}},
  \bibinfo{author}{\bibfnamefont{A.}~\bibnamefont{Osterloh}}, \bibnamefont{and}
  \bibinfo{author}{\bibfnamefont{V.}~\bibnamefont{Vedral}},
  \bibinfo{journal}{Rev. Mod. Phys.} \textbf{\bibinfo{volume}{80}},
  \bibinfo{pages}{517} (\bibinfo{year}{2008}).

\bibitem[{\citenamefont{G\"uhne and T\'oth}(2009)}]{guhne09}
\bibinfo{author}{\bibfnamefont{O.}~\bibnamefont{G\"uhne}} \bibnamefont{and}
  \bibinfo{author}{\bibfnamefont{G.}~\bibnamefont{T\'oth}},
  \bibinfo{journal}{Phys. Rep.} \textbf{\bibinfo{volume}{474}},
  \bibinfo{pages}{1} (\bibinfo{year}{2009}).

\bibitem[{\citenamefont{Wie\'sniak et~al.}(2005)\citenamefont{Wie\'sniak,
  Vedral, and Brukner}}]{wiesniak05}
\bibinfo{author}{\bibfnamefont{M.}~\bibnamefont{Wie\'sniak}},
  \bibinfo{author}{\bibfnamefont{V.}~\bibnamefont{Vedral}}, \bibnamefont{and}
  \bibinfo{author}{\bibfnamefont{C.}~\bibnamefont{Brukner}},
  \bibinfo{journal}{New J. Phys.} \textbf{\bibinfo{volume}{7}},
  \bibinfo{pages}{258} (\bibinfo{year}{2005}).

\bibitem[{\citenamefont{Ghosh et~al.}(2003)\citenamefont{Ghosh, Rosenbaum,
  Aeppli, and Coppersmith}}]{ghosh}
\bibinfo{author}{\bibfnamefont{S.}~\bibnamefont{Ghosh}},
  \bibinfo{author}{\bibfnamefont{T.~F.} \bibnamefont{Rosenbaum}},
  \bibinfo{author}{\bibfnamefont{G.}~\bibnamefont{Aeppli}}, \bibnamefont{and}
  \bibinfo{author}{\bibfnamefont{S.~N.} \bibnamefont{Coppersmith}},
  \bibinfo{journal}{Nature} \textbf{\bibinfo{volume}{425}}, \bibinfo{pages}{48}
  (\bibinfo{year}{2003}).

\bibitem[{\citenamefont{Brukner et~al.}(2006)\citenamefont{Brukner, Vedral, and
  Zeilinger}}]{brukner06}
\bibinfo{author}{\bibfnamefont{C.}~\bibnamefont{Brukner}},
  \bibinfo{author}{\bibfnamefont{V.}~\bibnamefont{Vedral}}, \bibnamefont{and}
  \bibinfo{author}{\bibfnamefont{A.}~\bibnamefont{Zeilinger}},
  \bibinfo{journal}{Phys. Rev. A} \textbf{\bibinfo{volume}{73}},
  \bibinfo{pages}{012110} (\bibinfo{year}{2006}).

\bibitem[{\citenamefont{Dowling et~al.}(2004)\citenamefont{Dowling, Doherty,
  and Bartlett}}]{PhysRevA.70.062113}
\bibinfo{author}{\bibfnamefont{M.~R.} \bibnamefont{Dowling}},
  \bibinfo{author}{\bibfnamefont{A.~C.} \bibnamefont{Doherty}},
  \bibnamefont{and} \bibinfo{author}{\bibfnamefont{S.~D.}
  \bibnamefont{Bartlett}}, \bibinfo{journal}{Phys. Rev. A}
  \textbf{\bibinfo{volume}{70}}, \bibinfo{pages}{062113}
  (\bibinfo{year}{2004}).

\bibitem[{\citenamefont{Brukner and Vedral}(2004)}]{brukner04}
\bibinfo{author}{\bibfnamefont{C.}~\bibnamefont{Brukner}} \bibnamefont{and}
  \bibinfo{author}{\bibfnamefont{V.}~\bibnamefont{Vedral}},
  \bibinfo{journal}{arXiv:quant-ph/0406040}  (\bibinfo{year}{2004}).

\bibitem[{\citenamefont{T\'oth}(2005)}]{toth05}
\bibinfo{author}{\bibfnamefont{G.}~\bibnamefont{T\'oth}},
  \bibinfo{journal}{Phys. Rev. A} \textbf{\bibinfo{volume}{71}},
  \bibinfo{pages}{010301} (\bibinfo{year}{2005}).

\bibitem[{\citenamefont{G\"uhne et~al.}(2005)\citenamefont{G\"uhne, T\'oth, and
  Briegel}}]{guhne05}
\bibinfo{author}{\bibfnamefont{O.}~\bibnamefont{G\"uhne}},
  \bibinfo{author}{\bibfnamefont{G.}~\bibnamefont{T\'oth}}, \bibnamefont{and}
  \bibinfo{author}{\bibfnamefont{H.~J.} \bibnamefont{Briegel}},
  \bibinfo{journal}{New J. Phys.} \textbf{\bibinfo{volume}{7}},
  \bibinfo{pages}{229} (\bibinfo{year}{2005}).

\bibitem[{\citenamefont{G\"uhne and T\'oth}(2006)}]{PhysRevA.73.052319}
\bibinfo{author}{\bibfnamefont{O.}~\bibnamefont{G\"uhne}} \bibnamefont{and}
  \bibinfo{author}{\bibfnamefont{G.}~\bibnamefont{T\'oth}},
  \bibinfo{journal}{Phys. Rev. A} \textbf{\bibinfo{volume}{73}},
  \bibinfo{pages}{052319} (\bibinfo{year}{2006}).

\bibitem[{\citenamefont{Duan}(2011)}]{PhysRevLett.107.180502}
\bibinfo{author}{\bibfnamefont{L.-M.} \bibnamefont{Duan}},
  \bibinfo{journal}{Phys. Rev. Lett.} \textbf{\bibinfo{volume}{107}},
  \bibinfo{pages}{180502} (\bibinfo{year}{2011}).

\bibitem[{\citenamefont{Sorensen et~al.}(2001)\citenamefont{Sorensen, Duan,
  Cirac, and Zoller}}]{sorensen}
\bibinfo{author}{\bibfnamefont{A.}~\bibnamefont{Sorensen}},
  \bibinfo{author}{\bibfnamefont{L.-M.} \bibnamefont{Duan}},
  \bibinfo{author}{\bibfnamefont{J.~I.} \bibnamefont{Cirac}}, \bibnamefont{and}
  \bibinfo{author}{\bibfnamefont{P.}~\bibnamefont{Zoller}},
  \bibinfo{journal}{Nature} \textbf{\bibinfo{volume}{409}}, \bibinfo{pages}{63}
  (\bibinfo{year}{2001}).

\bibitem[{\citenamefont{Korbicz et~al.}(2005)\citenamefont{Korbicz, Cirac, and
  Lewenstein}}]{PhysRevLett.95.120502}
\bibinfo{author}{\bibfnamefont{J.~K.} \bibnamefont{Korbicz}},
  \bibinfo{author}{\bibfnamefont{J.~I.} \bibnamefont{Cirac}}, \bibnamefont{and}
  \bibinfo{author}{\bibfnamefont{M.}~\bibnamefont{Lewenstein}},
  \bibinfo{journal}{Phys. Rev. Lett.} \textbf{\bibinfo{volume}{95}},
  \bibinfo{pages}{120502} (\bibinfo{year}{2005}).

\bibitem[{\citenamefont{Vitagliano et~al.}(2011)\citenamefont{Vitagliano,
  Hyllus, Egusquiza, and T\'oth}}]{PhysRevLett.107.240502}
\bibinfo{author}{\bibfnamefont{G.}~\bibnamefont{Vitagliano}},
  \bibinfo{author}{\bibfnamefont{P.}~\bibnamefont{Hyllus}},
  \bibinfo{author}{\bibfnamefont{I.~L.} \bibnamefont{Egusquiza}},
  \bibnamefont{and} \bibinfo{author}{\bibfnamefont{G.}~\bibnamefont{T\'oth}},
  \bibinfo{journal}{Phys. Rev. Lett.} \textbf{\bibinfo{volume}{107}},
  \bibinfo{pages}{240502} (\bibinfo{year}{2011}).

\bibitem[{\citenamefont{Bru\ss{} et~al.}(2005)\citenamefont{Bru\ss{}, Datta,
  Ekert, Kwek, and Macchiavello}}]{brus}
\bibinfo{author}{\bibfnamefont{D.}~\bibnamefont{Bru\ss{}}},
  \bibinfo{author}{\bibfnamefont{N.}~\bibnamefont{Datta}},
  \bibinfo{author}{\bibfnamefont{A.}~\bibnamefont{Ekert}},
  \bibinfo{author}{\bibfnamefont{L.~C.} \bibnamefont{Kwek}}, \bibnamefont{and}
  \bibinfo{author}{\bibfnamefont{C.}~\bibnamefont{Macchiavello}},
  \bibinfo{journal}{Phys. Rev. A} \textbf{\bibinfo{volume}{72}},
  \bibinfo{pages}{014301} (\bibinfo{year}{2005}).

\bibitem[{\citenamefont{Wang}(2002)}]{wangp}
\bibinfo{author}{\bibfnamefont{X.}~\bibnamefont{Wang}}, \bibinfo{journal}{Phys.
  Rev. A} \textbf{\bibinfo{volume}{66}}, \bibinfo{pages}{044305}
  (\bibinfo{year}{2002}).

\bibitem[{\citenamefont{Candini et~al.}(2010)\citenamefont{Candini, Lorusso,
  Troiani, Ghirri, Carretta, Santini, Amoretti, Muryn, Tuna, Timco
  et~al.}}]{candini10}
\bibinfo{author}{\bibfnamefont{A.}~\bibnamefont{Candini}},
  \bibinfo{author}{\bibfnamefont{G.}~\bibnamefont{Lorusso}},
  \bibinfo{author}{\bibfnamefont{F.}~\bibnamefont{Troiani}},
  \bibinfo{author}{\bibfnamefont{A.}~\bibnamefont{Ghirri}},
  \bibinfo{author}{\bibfnamefont{S.}~\bibnamefont{Carretta}},
  \bibinfo{author}{\bibfnamefont{P.}~\bibnamefont{Santini}},
  \bibinfo{author}{\bibfnamefont{G.}~\bibnamefont{Amoretti}},
  \bibinfo{author}{\bibfnamefont{C.}~\bibnamefont{Muryn}},
  \bibinfo{author}{\bibfnamefont{F.}~\bibnamefont{Tuna}},
  \bibinfo{author}{\bibfnamefont{G.}~\bibnamefont{Timco}},
  \bibnamefont{et~al.}, \bibinfo{journal}{Phys. Rev. Lett.}
  \textbf{\bibinfo{volume}{104}}, \bibinfo{pages}{037203}
  (\bibinfo{year}{2010}).

\bibitem[{\citenamefont{Troiani}(2011)}]{troiani11}
\bibinfo{author}{\bibfnamefont{F.}~\bibnamefont{Troiani}},
  \bibinfo{journal}{Phys. Rev. A} \textbf{\bibinfo{volume}{83}},
  \bibinfo{pages}{022324} (\bibinfo{year}{2011}).

\bibitem[{\citenamefont{Troiani et~al.}(2005)\citenamefont{Troiani, Ghirri,
  Affronte, Carretta, Santini, Amoretti, Piligkos, Timco, and
  Winpenny}}]{troiani05}
\bibinfo{author}{\bibfnamefont{F.}~\bibnamefont{Troiani}},
  \bibinfo{author}{\bibfnamefont{A.}~\bibnamefont{Ghirri}},
  \bibinfo{author}{\bibfnamefont{M.}~\bibnamefont{Affronte}},
  \bibinfo{author}{\bibfnamefont{S.}~\bibnamefont{Carretta}},
  \bibinfo{author}{\bibfnamefont{P.}~\bibnamefont{Santini}},
  \bibinfo{author}{\bibfnamefont{G.}~\bibnamefont{Amoretti}},
  \bibinfo{author}{\bibfnamefont{S.}~\bibnamefont{Piligkos}},
  \bibinfo{author}{\bibfnamefont{G.}~\bibnamefont{Timco}}, \bibnamefont{and}
  \bibinfo{author}{\bibfnamefont{R.~E.~P.} \bibnamefont{Winpenny}},
  \bibinfo{journal}{Phys. Rev. Lett.} \textbf{\bibinfo{volume}{94}},
  \bibinfo{pages}{207208} (\bibinfo{year}{2005}).

\bibitem[{\citenamefont{Trif et~al.}(2008)\citenamefont{Trif, Troiani,
  Stepanenko, and Loss}}]{trif08}
\bibinfo{author}{\bibfnamefont{M.}~\bibnamefont{Trif}},
  \bibinfo{author}{\bibfnamefont{F.}~\bibnamefont{Troiani}},
  \bibinfo{author}{\bibfnamefont{D.}~\bibnamefont{Stepanenko}},
  \bibnamefont{and} \bibinfo{author}{\bibfnamefont{D.}~\bibnamefont{Loss}},
  \bibinfo{journal}{Phys. Rev. Lett.} \textbf{\bibinfo{volume}{101}},
  \bibinfo{pages}{217201} (\bibinfo{year}{2008}).

\bibitem[{\citenamefont{Tsukerblat}(1994)}]{tsukerblat}
\bibinfo{author}{\bibfnamefont{B.}~\bibnamefont{Tsukerblat}},
  \emph{\bibinfo{title}{Group Theory in Chemistry and Spectroscopy}}
  (\bibinfo{publisher}{Academic Press, New York}, \bibinfo{year}{1994}).

\bibitem[{\citenamefont{Atkinson}(1989)}]{atkinson}
\bibinfo{author}{\bibfnamefont{K.~E.} \bibnamefont{Atkinson}},
  \emph{\bibinfo{title}{An introduction to numerical analysis}}
  (\bibinfo{publisher}{John Wiley \& Sons}, \bibinfo{year}{1989}).

\bibitem[{\citenamefont{Auerbach}(1994)}]{auerbach}
\bibinfo{author}{\bibfnamefont{A.}~\bibnamefont{Auerbach}},
  \emph{\bibinfo{title}{Interacting electrons and quantum magnetism}}
  (\bibinfo{publisher}{Springer}, \bibinfo{year}{1994}).

\end{thebibliography}
\end{document}